\documentclass[aps, prx, reprint, superscriptaddress]{revtex4-1}

\pdfoutput=1 
\usepackage{graphicx}
\usepackage{color,amsmath}
\usepackage[normalem]{ulem}

\usepackage[colorlinks,linkcolor=blue,citecolor=blue,urlcolor=blue]{hyperref}
\usepackage{siunitx}
\usepackage{comment}
\usepackage{layouts}

\begin{document}
\title{Few-electron Single and Double Quantum Dots \\in an InAs Two-Dimensional Electron Gas}

\author{Christopher Mittag}
\email{mittag@phys.ethz.ch}
\affiliation{Solid State Physics Laboratory, Department of Physics, ETH Zurich, 8093 Zurich, Switzerland}

\author{Jonne V. Koski}
\affiliation{Solid State Physics Laboratory, Department of Physics, ETH Zurich, 8093 Zurich, Switzerland}

\author{Matija Karalic}
\affiliation{Solid State Physics Laboratory, Department of Physics, ETH Zurich, 8093 Zurich, Switzerland}


\author{Candice Thomas}
\affiliation{Microsoft Station Q Purdue and Department of Physics and Astronomy, Purdue University, West Lafayette, Indiana 47907, USA}
\affiliation{Birck Nanotechnology Center, Purdue University, West Lafayette, Indiana 47907, USA}

\author{Aymeric Tuaz}
\affiliation{Microsoft Station Q Purdue and Department of Physics and Astronomy, Purdue University, West Lafayette, Indiana 47907, USA}
\affiliation{Birck Nanotechnology Center, Purdue University, West Lafayette, Indiana 47907, USA}

\author{Anthony T. Hatke}
\affiliation{Microsoft Station Q Purdue and Department of Physics and Astronomy, Purdue University, West Lafayette, Indiana 47907, USA}
\affiliation{Birck Nanotechnology Center, Purdue University, West Lafayette, Indiana 47907, USA}

\author{Geoffrey C. Gardner}
\affiliation{Microsoft Station Q Purdue and Department of Physics and Astronomy, Purdue University, West Lafayette, Indiana 47907, USA}
\affiliation{Birck Nanotechnology Center, Purdue University, West Lafayette, Indiana 47907, USA}

\author{Michael J. Manfra}
\affiliation{Microsoft Station Q Purdue and Department of Physics and Astronomy, Purdue University, West Lafayette, Indiana 47907, USA}
\affiliation{Birck Nanotechnology Center, Purdue University, West Lafayette, Indiana 47907, USA}

\author{Jeroen Danon}
\affiliation{Center for Quantum Spintronics, Department of Physics, Norwegian University of Science and Technology, NO-7491 Trondheim, Norway}

\author{Thomas Ihn}
\affiliation{Solid State Physics Laboratory, Department of Physics, ETH Zurich, 8093 Zurich, Switzerland}

\author{Klaus Ensslin}
\affiliation{Solid State Physics Laboratory, Department of Physics, ETH Zurich, 8093 Zurich, Switzerland}

\date{\today}

\begin{abstract}
Most proof-of-principle experiments for spin qubits have been performed using GaAs-based quantum dots because of the excellent control they offer over tunneling barriers and the orbital and spin degrees of freedom.
Here, we present the first realization of high-quality single and double quantum dots hosted in an InAs two-dimensional electron gas (2DEG),
demonstrating accurate control down to the few-electron regime, where we observe a clear Kondo effect and singlet-triplet spin blockade.
We measure an electronic $g$-factor of $16$ and a typical magnitude of the random hyperfine fields on the dots of $\sim \SI{0.6}{\milli\tesla}$.
We estimate the spin-orbit length in the system to be $\sim 5$--$\SI{10}{\micro\meter}$, which is almost two orders of magnitude longer than typically measured in InAs nanostructures, achieved by a very symmetric design of the quantum well.
These favorable properties put the InAs 2DEG on the map as a compelling host for studying fundamental aspects of spin qubits.
Furthermore, having weak spin-orbit coupling in a material with a large Rashba coefficient potentially opens up avenues for engineering structures with spin-orbit coupling that can be controlled locally in space and/or time.
\end{abstract}

\maketitle
\section{Introduction}

Electron spins hosted in semiconductor quantum dots are a compelling platform for quantum information processing\,\cite{loss_quantum_1998}, and two-dimensional electron gases (2DEGs) formed in GaAs/AlGaAs heterostructures have been the workhorse of this field for many years.
Due to the excellent gating technologies in this material system, few-electron quantum dots can be routinely realized with occupation control down to the last electron, promising in principle good scalability that is compatible with standard fabrication techniques.
The main obstacle to further progress in the field of GaAs-based spin qubits is the fast decoherence caused by the hyperfine coupling of the electron spins to the randomly fluctuating nuclear spins in the host material\,\cite{klg,frank:nature,Petta2005,Medford2013}.
The fact that this problem is intrinsic for GaAs was one of the reasons for a recent shift of focus toward Si- and Ge-based quantum dots, where the most abundant isotopes have zero nuclear spin.
Although progress has been impressive in recent years\,\cite{Yoneda2018,Huang2018c,Petit2019,Yang2020,Watzinger2018,Froning2020,Hendrickx2020a}, electrons in these materials have an additional valley degree of freedom which is difficult to control and can interfere with qubit operation\,\cite{Zwanenburg2013}.

An attractive alternative host system for studying fundamental aspects of spin qubits could be the InAs-based 2DEG.
Although both In and As carry non-zero nuclear spin, the bulk electronic $g$-factor in InAs is $\sim$~30 times larger than in GaAs, yielding much smaller effective nuclear fields for comparable hyperfine coupling and dot dimensions.
Apart from that, the small effective mass $m^* \approx 0.023\,m_0$ in the conduction band of InAs eases the demands on lithographic precision, coming with the additional advantage that a larger dot size reduces the r.m.s.\ value of the random nuclear fields even further.

The reason why the InAs-based 2DEG has not yet become a leading platform for spin-qubit implementations is twofold:
(i) The sidewalls of etched InAs structures usually contain an accumulation layer of electrons, likely caused by Fermi level pinning\,\cite{nichele_edge_2016,nguyen_decoupling_2016,mueller_edge_2017,mittag_passivation_2017}.
The traditional mesa etch technique commonly used in GaAs can thus not be applied, as the trivial edge states always short the electron gas underneath the gates traversing the mesa.
(ii) InAs has a large Rashba spin-orbit coefficient\,\cite{winkler_spin-orbit_2003,grundler_large_2000}, which can leverage small asymmetries in the confining potential into strong in-plane spin-orbit coupling.
This in fact resulted in a surge in interest in InAs-based devices, after it was pointed out that strong spin-orbit interaction in combination with proximity-induced superconductivity could result in a topological state hosting zero-energy Majorana modes at its boundaries\,\cite{kitaev_unpaired_2001,Sau2010a,lutchyn_majorana_2010,oreg_helical_2010}, which potentially could be used for topologically protected quantum computation\,\cite{PhysRevLett.86.268,RevModPhys.80.1083}.
Most experiments in this field rely on wire-like devices, since they allow superconductors to be more easily coupled to or even epitaxially grown on them\,\cite{mourik_signatures_2012,Deng2016a,Zhang}, but a planar setup would enable the application of top-down technology for the fabrication of complex quantum circuits\,\cite{Hell2017a}.
In the context of spin qubits, however, strong spin-orbit coupling is not necessarily beneficial: Although it does allow for fast electron dipole spin resonance, in general it reduces qubit coherence and compromises qubit control\,\cite{fasth_direct_2007,pfund_spin-state_2007,nadj-perge_disentangling_2010,nadj-perge_spinorbit_2010}.

Here, we present the first realization of high-quality single and double quantum dots using a split-gate technology on a planar InAs quantum well, demonstrating electron occupation control down to the last electron and a Kondo effect and spin blockade with qualities comparable to the best GaAs-based samples.
We circumvented the issue of edge conductance by using multiple layers of electrostatic gates\,\cite{mittag_edgeless_2018} and we suppressed the Rashba spin-orbit interaction by covering the InAs well with epitaxial AlGaSb barriers, yielding a very symmetric electric field across the quantum well\,\footnote{In experiments searching for possible beating patterns in Shubnikov-de Haas oscillations signaling spin-orbit split bands, a negative result was reported many years ago for quite similar samples as investigated in our present work\,\cite{brosig_zero-field_1999}.}.
We report a $g$-factor of $g \approx 16$ and an r.m.s.\ value of $\sim \SI{0.6}{\milli\tesla}$ for the random hyperfine fields on the dots, which is roughly an order of magnitude smaller than the typical fields in GaAs-based dots\,\cite{koppens_control_2005,jouravlev:prl}.
We further estimated the spin-orbit length $l_{\rm so}$ in our 2DEG in two independent ways showing that $l_{\rm so} \sim 5$--$\SI{10}{\micro\meter}$, which is of the same order of magnitude as in typical GaAs devices\,\cite{ronaldrev}.
This is unusually long for a lower-dimensional InAs-based structure and indicates that the Rashba contribution is strongly suppressed.

The implications for the field of quantum technologies are manifold.
First of all, we put the InAs 2DEG on the map as a new compelling platform for quantum-dot-based spin-qubit implementations, having the aforementioned benefits of a larger $g$-factor and smaller effective mass as compared to the commonly used GaAs.
The discovery that our symmetric heterostructure design indeed leads to an almost complete suppression of the Rashba effect removes most of the spin-orbit related decoherence and spin-mixing mechanisms, which in practice were considered to be intrinsic features in InAs-based devices.

Secondly, having a 2DEG with weak spin-orbit coupling in a material with a large Rashba coefficient presents conceptually a plethora of new possibilities.
In such a material gate-induced electric fields are capable of generating strong spin-orbit interaction locally in space and/or time.
In the context of spin qubits, one could thus imagine using a stacked-gate technology to switch on and off spin-orbit interaction locally on a quantum dot:
Switching it on allows for fast spin manipulation and switching it off could result in a long coherence time, thereby combining the best of the two worlds of strong and weak spin-orbit coupling.
Furthermore, adding the ingredient of induced superconductivity to a 2DEG with a gate-designable spin-orbit landscape could yield a convenient platform for creating and braiding Majorana bound states.

The rest of this manuscript is outlined as follows.
We begin by introducing the details of the heterostructure, the device, and its fabrication.
In the next section, we form and characterize a single few-electron quantum dot.
In an in-plane magnetic field we identify the spin of electronic states from the energy shift of the resonances visible in finite-bias spectroscopy and analyze the spin-orbit interaction based on the crossing of spin singlet and spin triplet states.
Completing the experiments on the single quantum dot, we increase the tunnel coupling to the leads and observe the Kondo effect in the single-electron regime; its dependence on bias voltage, magnetic field, and temperature are analyzed.
The next section comprises experiments on the coupling of two neighboring quantum dots to form a double quantum dot.
Singlet-triplet spin blockade by Pauli exclusion is observed and characterized.
Studying the leakage current at strong interdot tunnel coupling allows us to estimate the strength of the spin-orbit interaction, showing that it is much weaker than reported for other InAs-based devices.
Finally, we reduce the interdot tunnel coupling and use the narrow resonance in the leakage current around zero magnetic field to estimate the typical magnitude of the nuclear fields on the two dots.

\section{Device Details}
\begin{figure}[]
	\includegraphics[width=\columnwidth]{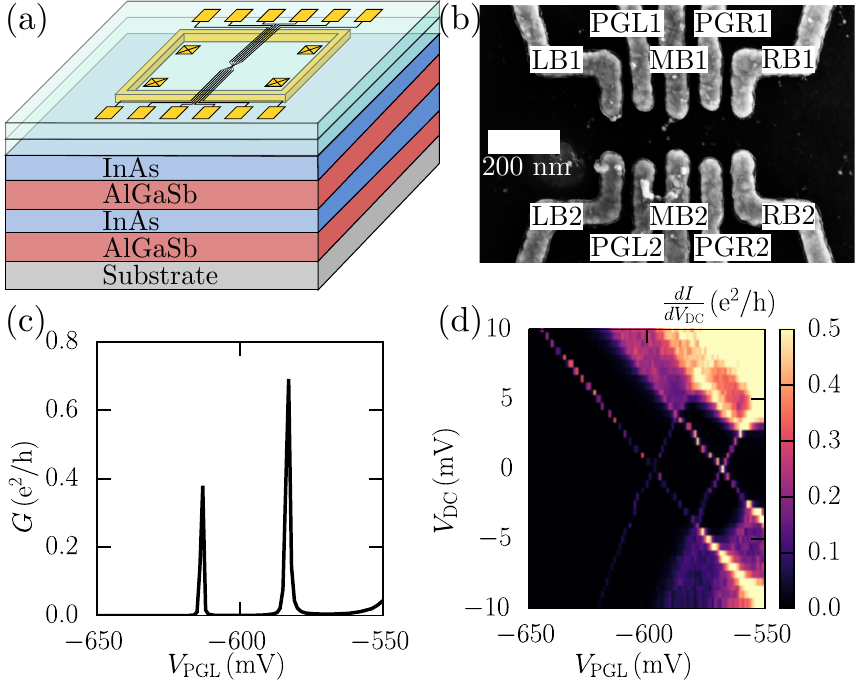}
	\caption{(a) Three-dimensional view of sample geometry with a cross-section of the heterostructure employed. The $\SI{24}{nm}$ wide InAs quantum well (lower blue layer) is sandwiched by two $\SI{20}{nm}$ AlGaSb barriers (red) and capped with a $\SI{1.5}{nm}$ InAs layer (upper blue layer). Multiple layers of gates separated by $\mathrm{Al}_{\mathrm{2}}\mathrm{O}_{\mathrm{3}}$ dielectric layers create a device used to form single and double quantum dots. (b) Scanning electron micrograph of the gate layout of a device similar to the one used in this study. (c) Coulomb resonances in the conductance $G$ through the left quantum dot as a function of $V_{\mathrm{PGL}}$. (d) Coulomb blockade diamond visible in the differential conductance as a function of $V_{\mathrm{PGL}}$ and $V_{\mathrm{DC}}$. Uninterrupted Coulomb resonances up to $V_{\mathrm{DC}}=\SI{\pm 10}{\milli\volt}$ and the absence of additional resonances indicate the loading of the first electron.} 
	\label{fig1}
\end{figure}

The heterostructure used in this experiment was grown by molecular beam epitaxy on an undoped GaSb substrate wafer in (100) crystal orientation. The main layers of importance for this work can be seen in the schematic cross-section in Fig.\,\ref{fig1}\,(a). They comprise a $\SI{24}{nm}$ wide InAs quantum well (lower blue layer in Fig.\,\ref{fig1}\,(a)) sandwiched between two  $\SI{20}{nm}$ wide $\mathrm{Al}_{0.8}\mathrm{Ga}_{0.2}\mathrm{Sb}$ barriers (red) and capped with a $\SI{1.5}{nm}$ InAs layer (upper blue layer). The quantum well hosts a 2DEG with a mobility $\mu=1.4\times10^6\,\SI{}{\centi\meter\squared\per\volt\per\second}$ at a density $n_{\mathrm{e}}=5.1\times 10^{11}\,\SI{}{\per\centi\meter\squared}$, measured at the temperature $T=\SI{1.5}{\kelvin}$.  Further details on hetereostructure growth and properties can be found in Ref.\,\cite{thomas_high-mobility_2018}. 

In order to fabricate the device, Ti/Ni/Au Ohmic contacts of thickness $(20/100/100)\,\SI{}{\nano\meter}$ are deposited in the first step. This is followed by the growth of $\SI{20}{nm}$ $\mathrm{Al}_{2}\mathrm{O}_{3}$ by atomic layer deposition at $\SI{150}{\celsius}$ as a first gate dielectric, on top of which the $(5/45)\,\SI{}{\nano\meter}$ (Ti/Au) rectangular frame gate is deposited. Subsequently, a second gate dielectric of $\SI{20}{nm}$ $\mathrm{Al}_{2}\mathrm{O}_{3}$ is deposited in analogy to the previous step. This enables the deposition of $(10/50)\,\SI{}{\nano\meter}$ (Ti/Au) gate leads and $(5/20)\,\SI{}{\nano\meter}$ (Ti/Au) fine gates defined by optical and electron beam lithography, respectively. In a last step, the contacts are released with a hydrofluoric acid etch of the dielectric layer in the contact areas and carefully contacted by glueing bond wires with conductive epoxy in order to prevent any leakage to the GaSb substrate wafer that could be caused by wire bonding. A schematic overview of this structure can be seen in Fig.\,\ref{fig1}\,(a). 

The scanning electron micrograph of in Fig.\,\ref{fig1}\,(b) shows the fine gate structure used to define single and double quantum dots of a sample similar to the one used in this study. The left quantum dot is defined by the left tunnel barriers (labeled LB1 and LB2), the left plunger gates (PGL1 and PGL2), and the middle tunnel barriers (MB1 and MB2). The latter gates form the right quantum dot together with the right plunger gates (PGR1 and PGR2) and the right tunnel barriers (RB1 and RB2). The measurements in this paper were conducted in a dilution refrigerator at the base temperature $T=\SI{57}{\milli\kelvin}$ using low-frequency AC lock-in techniques. The frame gate is kept at a negative voltage such that it depletes the 2DEG underneath and electronic transport is measured strictly in the inner part of the electron gas across the quantum dot structure. This is necessary to avoid spurious effects of conduction along the sample edges that were found in similar InAs devices\,\cite{nichele_edge_2016,nguyen_decoupling_2016,mueller_edge_2017,mittag_edgeless_2018}.

\section{Single Quantum Dot}
In the first experiment, we utilize the three leftmost gate pairs of Fig.\,\ref{fig1}\,(b) and set a negative potential to them to define the left quantum dot. In the conductance $G$ measured across the quantum dot as a function of $V_{\mathrm{PGL}}$, the voltage applied to the left plunger gates, we observe narrow Coulomb resonances. Two of these are shown in Fig.\,\ref{fig1}\,(c) and each indicates the change of the dot occupancy by one electron. We have tuned the system into a state where we presume these to be the last two electrons on the dot. As this sample does not have a nearby charge detector\,\cite{schleser_time-resolved_2004,vandersypen_real-time_2004} we present multiple experiments in the following which all consistently confirm that this is the case. One of these experiments is displayed in Fig.\,\ref{fig1}\,(d) where we record the differential conductance as a function of $V_{\mathrm{PGL}}$ and a DC source-drain bias voltage $V_{\mathrm{DC}}$ up to $\pm\SI{10}{\milli\volt}$, much larger than the charging energy of the quantum dot. Past the last Coulomb blockade diamond, no additional resonances appear while the resonances that indicate the charging of the dot with the first electron remain visible. This implies that the lack of additional resonances is not caused by the closing tunnel barriers. 

\subsection{Excited State Spectroscopy}
\begin{figure}[]
	\includegraphics[width=\columnwidth]{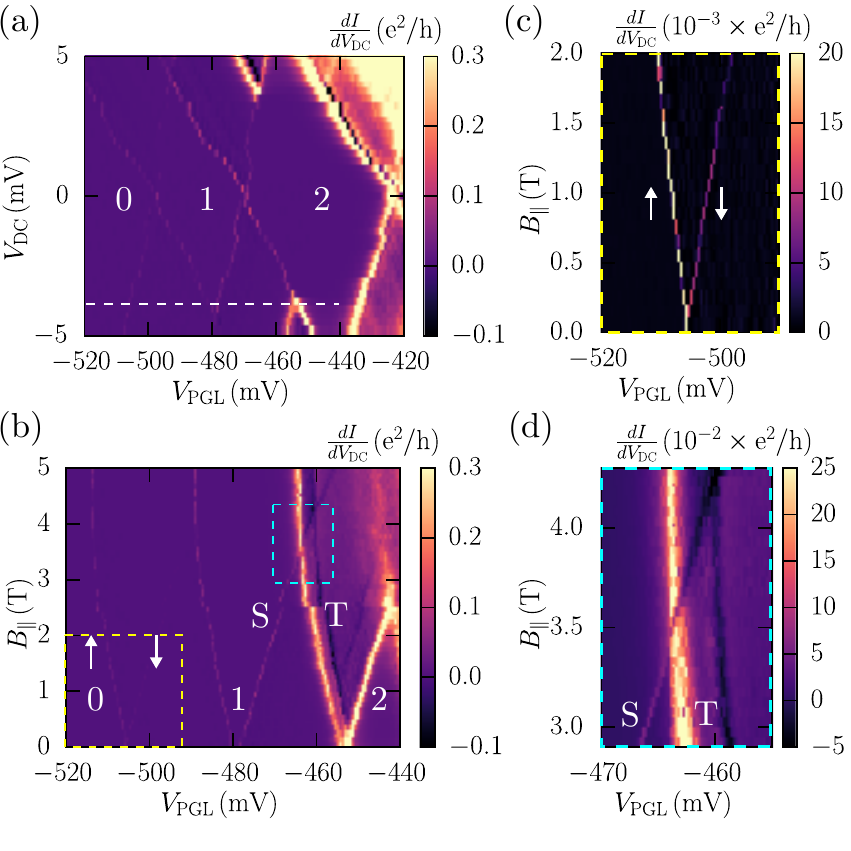}
	\caption{(a) Coulomb blockade diamonds corresponding to the loading of the first two electrons. A resonance corresponding to the transition to an excited state branches off the two-electron Coulomb blockade diamond at $V_{\mathrm{DC}}=\SI{-3.9}{\milli\volt}$. (b) Magnetic field dependence of the resonances along the white dashed line in (a). (c) Zoom into the yellow box in (b). A linear Zeeman splitting of the resonances corresponding to the loading of the first electron is visible. (d) Zoom into the blue box in (b) where a direct crossing of the resonances corresponding to the spin singlet and a spin triplet state is observed.}
	\label{fig2}
\end{figure}

For the next measurement, we have tuned the quantum dot such that we observe the Coulomb blockade diamonds of the first two electrons. These are shown in Fig.\,\ref{fig2}\,(a) where we again record the differential conductance as a function of $V_{\mathrm{PGL}}$ and $V_{\mathrm{DC}}$. Integer numbers indicate the dot occupancy. At a bias voltage $V_{\mathrm{DC}}=\pm\SI{3.9}{\milli\volt}$, we find resonances of excited states branching off the two-electron Coulomb blockade diamond. In two-electron quantum dots in GaAs, the ground state is usually believed to be the spin-singlet state, whereas this lowest excitation is associated with the three-fold degenerate spin-triplet state\,\cite{ellenberger_excitation_2006}.

In order to further elucidate the spin of these excited states, as well as to corroborate our hypothesis of the occupancy of this few-electron quantum dot we employ a parallel magnetic field while varying the plunger gate voltage along the white dashed line in Fig.\,\ref{fig2}\,(a) at a finite bias voltage  $V_{\mathrm{DC}}=\SI{-3.9}{\milli\volt}$. The resulting data of the differential conductance as a function of $V_{\mathrm{PGL}}$ and the magnetic field $B_{\parallel}$ is shown in Fig.\,\ref{fig2}\,(b). Here we note a linear Zeeman splitting of all resonances. To better visualize the splitting of the first visible resonance, we show in Fig.\,\ref{fig2}\,(c) a zoom into the yellow dashed box of Fig.\,\ref{fig2}\,(b). As expected for the first electron, we recognize a linear splitting into two resonances for the two spin projections. From the magnitude of the splitting, we extract an effective $g$-factor of $\vert g^*\vert =16.4$ which is comparable to the bulk value expected for InAs\,\cite{konopka_conduction_1967,pidgeon_interband_1967}.

\subsection{Singlet-Triplet Crossing}
One of the main results of this work is found inside the blue dashed box in Fig.\,\ref{fig2}\,(b), where the resonances marked `S' and `T' cross.
Resonance S appears where the transition energy ${\uparrow} \to S$ equals the chemical potential $\mu_{\rm S}$ of the source lead, and T where the energy of both ${\uparrow} \to T_+$ and ${\downarrow} \to T_0$ match $\mu_{\rm S}$.
Thus, at the point where S and T cross the two-electron $S$ and $T_+$ states are degenerate and past this crossing the ground state of the two-electron system turns from a spin singlet to a spin triplet\,\footnote{The fact that resonance S changes sign when it crosses T signals that the actual rates corresponding to all transitions contributing to electron transport are significantly different, most likely due to differences in the orbital structure of the states involved and/or asymmetric coupling to the source and drain leads.}.
This region of proximity between these two states has been thoroughly investigated in quantum dots formed in InAs nanowires, as the magnitude of the anticrossing observed there is indicative of the strength of the spin-orbit interaction in the quantum dot\,\cite{fasth_direct_2007,pfund_spin-state_2007}. The data in this region in our quantum dot defined in a 2DEG can be seen in Fig.\,\ref{fig2}\,(d). In contrast to the quantum dots defined in nanowires, we cannot detect a clear spin-orbit mediated anticrossing of singlet and triplet states within our experimental resolution. Converting $V_{\rm PGL}$ to an energy scale using the Zeeman splitting between the single-electron spin-up and -down resonances that are visible on the left side of Fig.\,\ref{fig2}\,(b), we estimate that the size of the anticrossing does not exceed $\sim 80~\mu$eV.
Using that the magnitude of this spin-orbit induced anticrossing is roughly $|g^*\mu_{\rm B}B_\parallel|r/\sqrt 2 l_{\rm so}$\,\cite{fasth_direct_2007}, where $r$ is the average distance between the two electrons occupying the dot, we can estimate the relevant spin-orbit length in the quantum dot to be $l_{\rm so} \gtrsim 0.9~\mu$m, assuming that $r \sim 30$~nm.

This corresponds to a rather weak spin-orbit coupling compared to other reported values for InAs-based systems\,\cite{grundler_large_2000,nitta_gate_1997,fasth_direct_2007,pfund_spin-state_2007}. In order to understand the relatively weak spin-orbit coupling we observe, we consider the various contributions separately: (i) The Dresselhaus contribution\,\cite{dresselhaus_spin-orbit_1955} to the spin-orbit interaction depends on crystal inversion asymmetry and its strength in InAs is comparable to GaAs\,\cite{winkler_spin-orbit_2003}. Due to the difference in effective mass, the associated spin-orbit length will be three times larger than in a comparable GaAs-based system and thus can be roughly estimated to be of the order of $3-30\,\SI{}{\micro\meter}$. The expected strength of the Dresselhaus contribution is thus not inconsistent with the estimate given above. (ii) The coefficient of the Rashba contribution, which determines the strength of the spin-orbit coupling arising from structural inversion asymmetry of the system, is much larger in InAs than in GaAs. Nevertheless, the Rashba-type spin-orbit interaction in the 2DEG does not only depend on this coefficient, but also on the electric field normal to the plane of the 2DEG. The large and tunable spin-orbit interaction that was attributed to InAs 2DEGs in previous experiments was achieved by employing very asymmetric quantum wells and strong electric fields perpendicular to the quantum well\,\cite{grundler_large_2000,nitta_gate_1997}. Here, on the other hand, the quantum well is wide and contained by symmetric barriers, presumably leading to a very symmetric wave function. In that case, the large Rashba coefficient can not be efficiently leveraged by an electric field and does therefore not lead to an overall strong spin-orbit interaction. This is consistent with a missing beating in the Shubnikov-de Haas oscillations observed both in our planar InAs electron gas and previously in similar heterostructures\,\cite{brosig_zero-field_1999}.

\subsection{Kondo Effect}
\begin{figure}[]
	\includegraphics[width=\columnwidth]{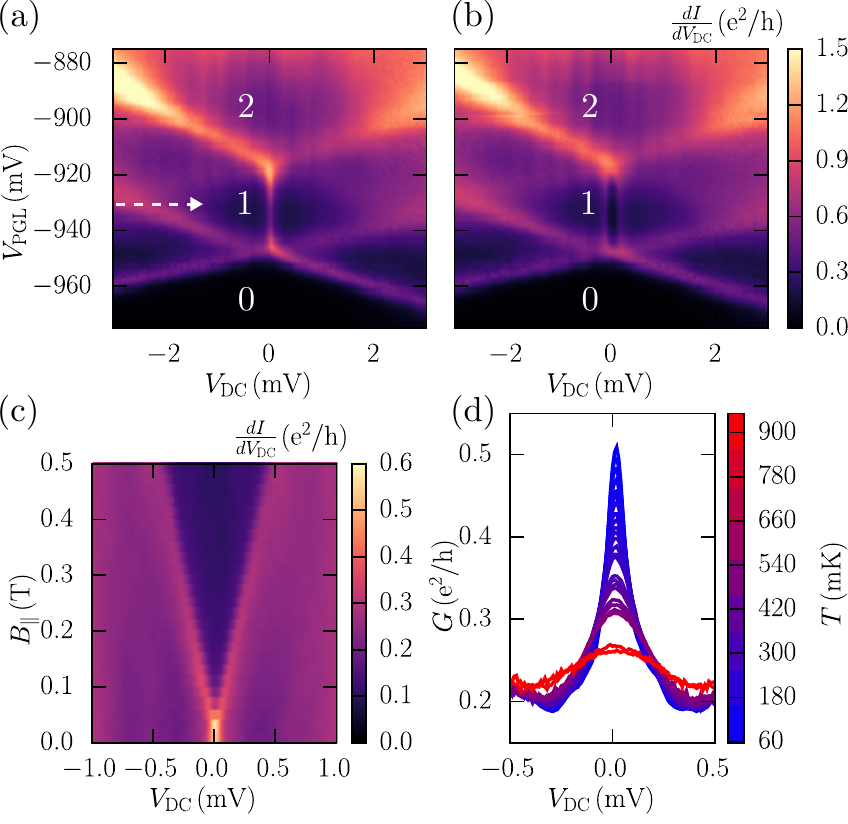}
	\caption{(a) Coulomb blockade diamond in the conduction of the left quantum dot at strong coupling to the leads as a function of $V_{\mathrm{DC}}$ and $V_{\mathrm{PGL}}$ at $B=0$. A clear zero-bias conductance peak indicates the observation of the Kondo effect. (b) Same as in (a) but at an external magnetic field of $B_{\parallel}=\SI{150}{\milli\tesla}$ causing a Zeeman splitting of the Kondo resonance. (c) Dependence of the Zeeman splitting of the Kondo resonance on $B_{\parallel}$ recorded at $V_{\mathrm{PGL}}=\SI{-933}{\milli\volt}$ marked by the dashed arrow in (a). (d) Temperature dependence of the peak conductance of the Kondo resonance.} 
	\label{fig3}
\end{figure}

Further investigating the quantum dot, we increase the tunnel coupling to the leads and observe the Kondo effect\,\cite{goldhaber-gordon_kondo_1998,cronenwett_tunable_1998}. This phenomenon leads to a zero-bias resonance when an unpaired spin on the quantum dot forms a many-body state with the electrons in the leads. In Fig.\,\ref{fig3}\,(a) we show a measurement of the differential conductance while varying $V_{\mathrm{DC}}$ and $V_{\mathrm{PGL}}$. The Coulomb blockade diamond for the first electron occupying the quantum dot features a resonance at zero bias indicative of the Kondo effect in our system. This observation is consistent with our picture of the system thus far as it means that an unpaired spin is occupying the quantum dot. Fig.\,\ref{fig3}\,(b) displays the same measurement at an external magnetic field $B_{\parallel}=\SI{150}{\milli\tesla}$. We observe a Zeeman split Kondo resonance within the Coulomb blockade diamond, as is theoretically expected.
We investigate this splitting as a function of  $B_{\parallel}$ in Fig.\,\ref{fig3}\,(c). At constant  $V_{\mathrm{PGL}}=\SI{-933}{\milli\volt}$, marked by the white dashed arrow in Fig.\,\ref{fig3}\,(a), the position in $V_{\mathrm{DC}}$ of the Kondo resonance in the differential conductance linearly splits with increasing $B_{\parallel}$. For the Kondo effect, we expect twice the usual Zeeman splitting, $\Delta E = 2\vert g \vert \mu_{\mathrm{b}}B_{\parallel}$\,\cite{kretinin_spin-$frac12$_2011}. This yields a $g$-factor of $\vert g \vert=16.3$, in good agreement with the value extracted from the excited state spectroscopy. The analysis of the Kondo effect is completed by an investigation of its temperature dependence shown in Fig.\,\ref{fig3}\,(d). There, we plot the Kondo resonance as a function of $V_{\mathrm{DC}}$ at different temperatures ranging from $T=\SI{57}{\milli\kelvin}$ to $T=\SI{952}{\milli\kelvin}$ according to the legend at the right hand side of the figure. A strong suppression of the peak conductance with increasing temperature can be observed. We fit this behavior to the phenomenological expression $G(T)=G_0\left[ 1+(T/T'_{\mathrm{K}})^{2}\right]^{-s}$\,\cite{kretinin_spin-$frac12$_2011}, where $T'_{\mathrm{K}}=T_{\mathrm{K}}/(2^{1/s}-1)^{1/2}$, the parameter $s=0.22$, and $T_{\mathrm{K}}$ is the Kondo temperature. This results in a Kondo temperature of $T_{\mathrm{K}}\approx\SI{800}{\milli\kelvin}$.

\section{Double Quantum Dot}
\subsection{Charge Stability Diagram}
\begin{figure}[]
	\includegraphics[width=\columnwidth]{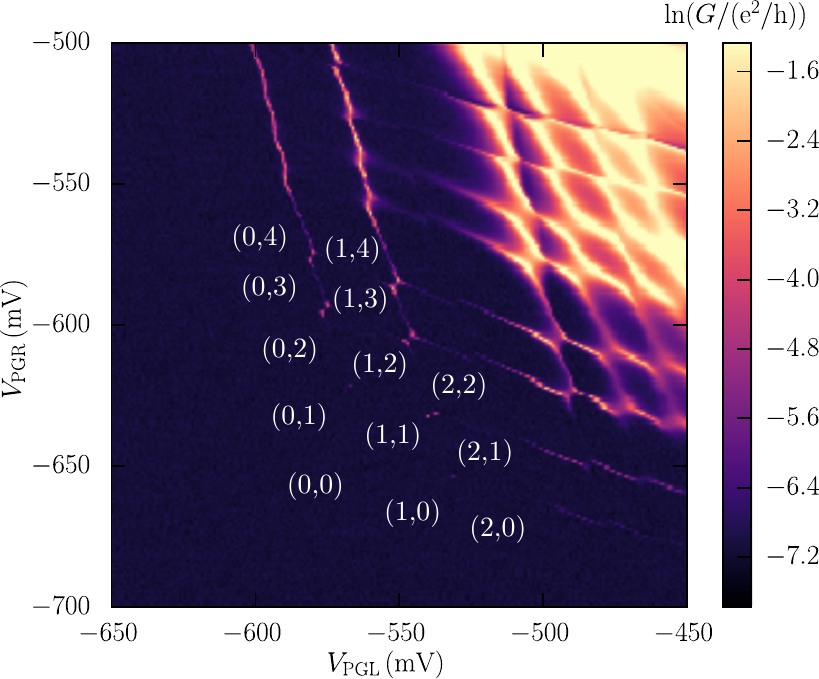}
	\caption{Charge stability diagram visible in the logarithm of the conductance $G$ of the double quantum dot system. Regions of stable charge occupancy are labeled with tuples $(N_{\mathrm{L}},N_{\mathrm{R}})$ where $N_{\mathrm{L}}$ ($N_{\mathrm{R}}$) indicates the occupancy of the left (right) quantum dot.}  
	\label{fig4}
\end{figure}

For the next set of experiments, we employ all the gates that can be seen in Fig.\,\ref{fig1}\,(b) to define a double quantum dot consisting of two tunnel coupled quantum dots in series. To characterize this system, we record a charge stability diagram. This is plotted in Fig.\,\ref{fig4} as the logarithm of the conductance as a function of the plunger gate voltages of the left quantum dot, $V_{\mathrm{PGL}}$, and of the right quantum dot, $V_{\mathrm{PGR}}$. The regions of stable charge occupancy are labelled with tuples $(N_{\mathrm{L}},N_{\mathrm{R}})$ representing the occupancy in the left and right quantum dot, respectively, and take the typical honeycomb shape that is expected from a double quantum dot. 

\subsection{Singlet-Triplet Spin Blockade}
\begin{figure}[]
	\includegraphics[width=\columnwidth]{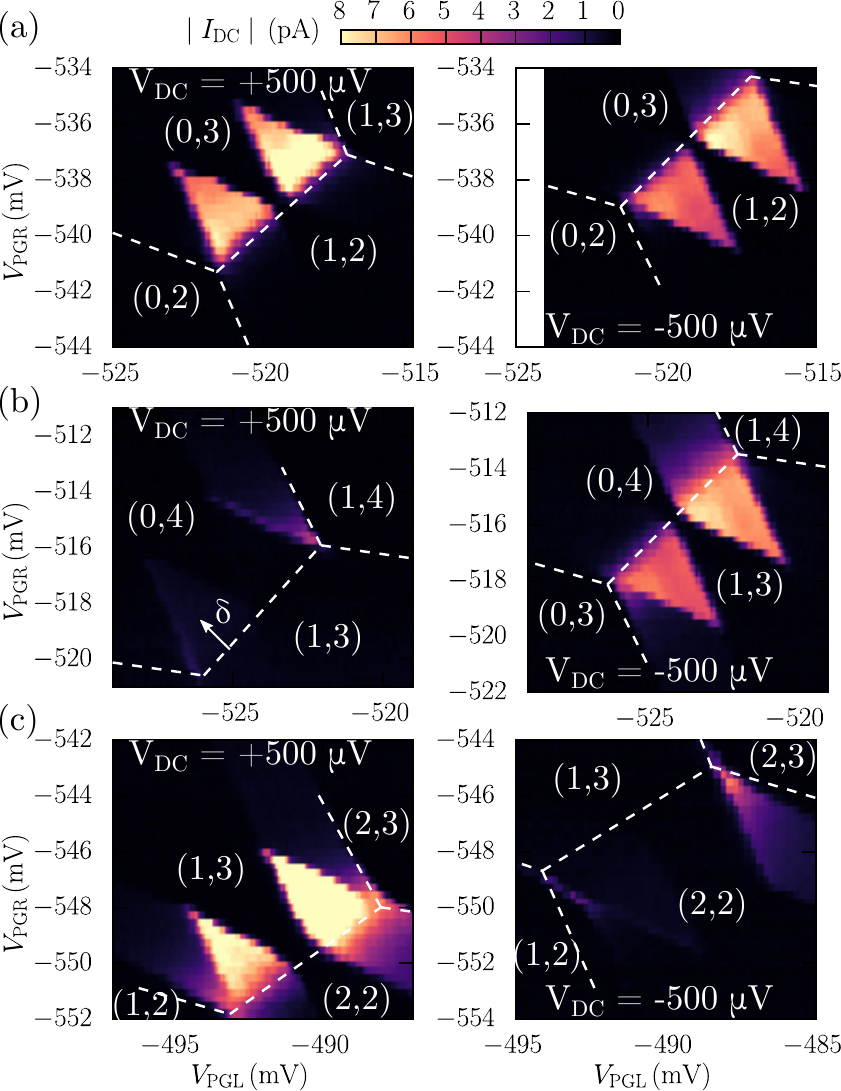}
	\caption{(a) Magnitude of the current $\vert I_{\mathrm{DC}}\vert$ through the double quantum dot system at finite bias of $V_{\mathrm{DC}}=\SI{+500}{\micro\volt}$ ($V_{\mathrm{DC}}=\SI{-500}{\micro\volt}$) in the left (right) panel. Electron transport involves the internal (1,2) to (0,3) (left) and (0,3) to (1,2) (right) transition. As expected, in both cases current can flow resulting in finite bias triangles. (b) $\vert I_{\mathrm{DC}}\vert$ involving the (1,3) to (0,4) and the (0,4) to (1,3) transitions in the left (right) panel. The triangles involving the transition from (1,3) to (0,4) occupancy at $V_{\mathrm{DC}}=\SI{+500}{\micro\volt}$ show an absence of current due to singlet-triplet spin blockade by Pauli exclusion. (c) $\vert I_{\mathrm{DC}}\vert$ involving the (2,2) to (1,3) and the (1,3) to (2,2) transitions in the left (right) panel. The transition from (1,3) to (2,2) occupancy at $V_{\mathrm{DC}}=\SI{-500}{\micro\volt}$ is blocked.}  
	\label{fig5}
\end{figure}

When applying a source-drain bias voltage, finite bias triangles of current flow are formed at the triple points where three regions of stable charge occupancy meet. In Fig.\,\ref{fig5} we show these triangles for three different sets of triple points, where we plot the magnitude of the current $|I_{\rm DC}|$ as a function of $V_{\rm PGL}$ and $V_{\rm PGR}$. The bias triangles involving the transition from (1,2) to (0,3) at $V_{\mathrm{DC}}=\SI{500}{\micro\volt}$ and those involving the reverse transition from (0,3) to (1,2) at $V_{\mathrm{DC}}=\SI{-500}{\micro\volt}$ can be seen in the left and right panel of Fig.\,\ref{fig5}\,(a). Regular and pronounced finite bias triangles are visible for both bias directions, as is expected since the charge transition (1,2) to (0,3) is not spin selective.
In Fig.\,\ref{fig5}\,(b) we investigate the border between the (0,4) and (1,3) regions, when one additional electron is in the double quantum dot system. A clear suppression of the current is visible in the left panel which involves the charge transition (1,3) to (0,4) as opposed to the right panel, where current involves the (0,4) to (1,3) transition. This can be explained by singlet-triplet spin blockade based on the Pauli exclusion principle\,\cite{ono_current_2002}: 

When the fourth electron tunnels onto the left dot, i.e., the (0,3) to (1,3) transition, there is no preferred direction of spin and the system can end up in any of the energetically allowed (1,3) spin states. Due to the Pauli principle, the only accessible (0,4) state is a spin singlet\,\footnote{In principle, due to intradot Coulomb interaction the four-electron ground state in a quantum dot can also be a spin triplet, depending on the shape and size of the dot. However, our field-dependent measurements of the current in the spin-blockade regime suggest that the (0,4) ground state is a singlet, see Sec. IV.C.} and the subsequent charge transition (1,3) to (0,4) is thus only allowed if the (1,3) state is in a singlet configuration. In the case a (1,3) spin triplet was formed, the electron becomes trapped in the left quantum dot and current flow ceases until a non-spin-conserving tunnel process takes place. Meanwhile, the system is said to be in spin blockade. The reversed bias involves the (0,4) to (1,3) transition, which can always take place since the (1,3) state has no preferred spin configuration. Therefore, there is no spin blockade observed in the right panel.

Similarly, at the border between the (2,2) and (1,3) regions, singlet-triplet spin blockade should occur when attempting to drive the system from (1,3) to (2,2) at a negative bias voltage. This is exactly what is observed in the right hand panel Fig.\,\ref{fig5}\,(c), where current is suppressed except at the outer edges where thermal broadening contributes to a finite current. At positive bias in the left panel we measure regular finite bias triangles since the transition from (2,2) to (1,3) is not spin selective. The occurrence of spin blockade is another experimental observation that corroborates our hypothesis of low occupancy of the quantum dot system, as it is only observed in the few-electron regime\,\cite{johnson_singlet-triplet_2005}.

\subsection{Spin Blockade at Strong Interdot Tunnel Coupling}
\begin{figure}[]
	\includegraphics[width=\columnwidth]{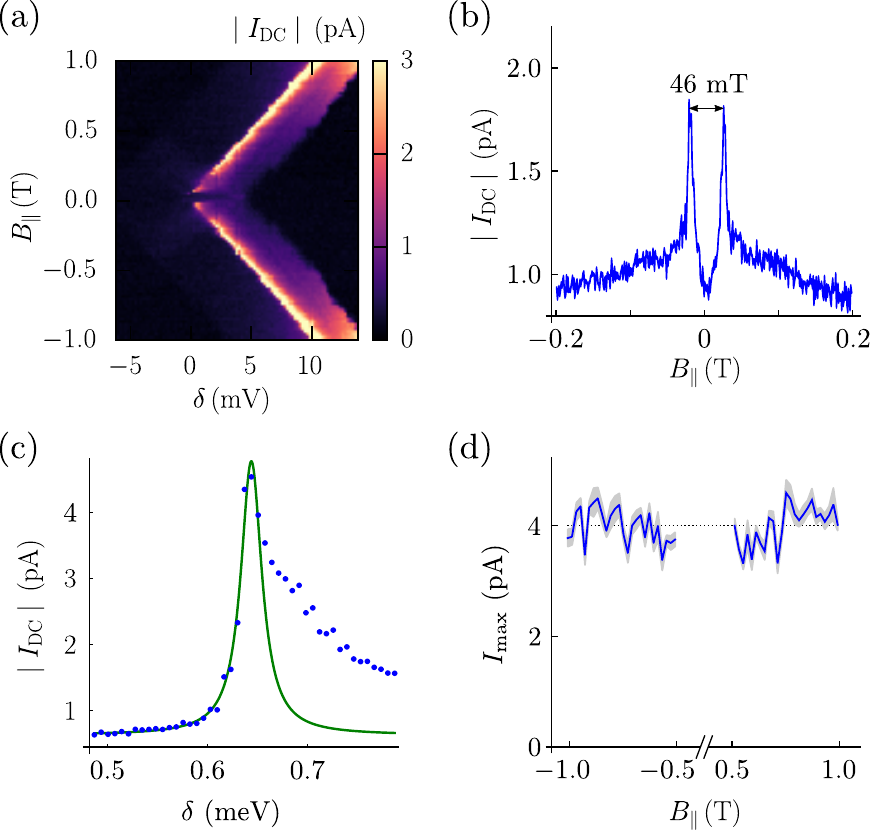}
	\caption{(a) Dependence of the magnitude of the leakage current $\vert I_{\mathrm{DC}}\vert$ through the spin blockade involving the (1,3) to (0,4) transition at strong interdot tunnel coupling on $\delta$ and $B_{\parallel}$. A dip around zero external magnetic field is visible in the data. (b) High-resolution trace of $\vert I_{\mathrm{DC}}\vert$ as a function of $B_{\parallel}$ around zero detuning. (c) $\delta$-dependent leakage current $\vert I_{\mathrm{DC}}\vert$ at $B_\parallel=\SI{700}{\milli\tesla}$ (blue) and a Lorentzian fit to the onset of the peak (green) disregarding the inelastic current tail. (d) Peak current values $I_{\rm max}$ for $|B_\parallel| > \SI{500}{\milli\tesla}$, where the grey shaded area indicates the magnitude of the error of each fit. Their field-independent height suggests spin-orbit coupling as the main mechanism responsible for spin-flip tunneling rather than hyperfine coupling.}  
	\label{fig6}
\end{figure}

We further characterize the spin blockade in our system, first at strong tunnel coupling between the two quantum dots. For this, we employ a parallel magnetic field $B_\parallel$ and change the detuning $\delta$ between the left and right dot, where $\delta$ is measured along the white arrow in Fig.\,\ref{fig5}\,(b) and is expressed in $\SI{}{\milli\electronvolt}$, where we used the extent of the bias triangles to find the lever arm. In Fig.\,\ref{fig6}\,(a) we plot the leakage current in the spin blockade regime involving the (1,3) to (0,4) transition, as a function of $B_\parallel$ and $\delta$. We will analyze these data and thereby try to extract information about the strength of the spin-orbit coupling.
The general structure of the $\delta$-dependent current, away from zero field, is a sharp resonance followed by a lower-current tail. The resonance is expected to appear where the (1,3) and (0,4) ground states align, and we attribute the tail at larger detuning to inelastic interdot tunneling where the excess energy of the (1,3) state is dissipated into the phonon bath. The current drops abruptly when $\delta$ is increased beyond the borders of the bias triangle, where the system enters Coulomb blockade, approximately $|eV_{\mathrm{DC}}|=\SI{0.5}{\milli\electronvolt}$ away from the baseline. The fact that the sharp resonances show a V-shape in the $(\delta,B_\parallel)$-plane is an indication that the (0,4) ground state is a spin singlet: Indeed, if the (0,4) ground state were a spin triplet, then the resonance between the (1,3) and (0,4) triplet ground states should occur around the same value of $\delta$ for all $B_\parallel$. Furthermore, from the slope of the resonances we find $|g^*| = 16.4$, which is in good agreement with our earlier estimates.

Extrapolating the two linear resonances to zero field allows us to identify the gate voltage settings that correspond to effective alignment of the energies of the (1,3) and (0,4) charge states. From a high-resolution trace along $B_\parallel$ very close to this point, shown in Fig.\,\ref{fig6}\,(b), we find a residual peak splitting of $\SI{46}{\milli\tesla}$, which corresponds to $\SI{44}{\micro\electronvolt}$, close to zero detuning. This provides a measure for $2t$ where $t$ is the coherent tunnel coupling strength between the (1,3) and (0,4) singlets\,\cite{koppens_control_2005}.

Next, we focus on the current as a function of $\delta$ for fields $|B_\parallel| > \SI{500}{\milli\tesla}$. At such large fields, the Zeeman splitting exceeds the size of the bias window $eV_{\rm DC}$ and we can thus safely assume that transport involves only ground states, i.e., the transport cycle is $D_\uparrow^{(0,3)} \to T_+^{(1,3)} \to S^{(0,4)} \to D_\uparrow^{(0,3)}$, where $D_\uparrow^{(0,3)}$ denotes the (0,3) charge state in a spin up configuration. In that simple situation, where only three states are involved, the resonant current is expected to have a Lorentzian line shape\,\cite{stoof_time-dependent_1996},
\begin{equation}
|I_{\rm DC}| = \frac{e t^2_{\rm sf}\Gamma_{\rm out}}{t^2_{\rm sf}(2+\Gamma_{\rm out}/\Gamma_{\rm in}) + (\hbar\Gamma_{\rm out})^2/4 + \delta^2},
\label{eq:two-level-resonance}
\end{equation}
where $\Gamma_{\rm out,in}$ denote the tunnel rates out of $S^{(0,4)}$ and into $T_+^{(1,3)}$, respectively. The coherent coupling between $T_+^{(1,3)}$ and $S^{(0,4)}$ involves a spin flip and its strength is denoted by $t_{\rm sf}$. Assuming that $\Gamma_{\rm out} \sim \Gamma_{\rm in}$ and anticipating that $t_{\rm sf} \ll \Gamma_{\rm out}$, Eq.~(\ref{eq:two-level-resonance}) reduces to a peak with a full width at half maximum of $\hbar \Gamma_{\rm out}$ and a peak value of $I_{\rm max} = 4et_{\rm sf}^2/\hbar^2\Gamma_{\rm out}$. Thus focusing on the sharp resonance in the data and disregarding the inelastic current tail, we fit the onset of each peak in the data in Fig.\,\ref{fig6}\,(c) to a Lorentzian (adding a constant background offset as fit parameter). The result of one such fit, for $B_\parallel=\SI{700}{\milli\tesla}$, is shown in Fig.\,\ref{fig6}\,(c) where the blue points present the data and the green curve the fitted Lorentzian, yielding $\hbar\Gamma_{\rm out} = \SI{25.5}{\micro\electronvolt}$ and $I_{\rm max} = \SI{4.14}{\pico\ampere}$ (which does not include the constant background). We repeat this procedure for all traces with $|B_\parallel| > \SI{500}{\milli\tesla}$, which results after averaging in $\hbar\Gamma_{\rm out} = 31 \pm 10\,\SI{}{\micro\electronvolt}$ and $I_{\rm max} =4.00\pm 0.34\,\SI{}{\pico\ampere}$. Using these values we find $t_{\rm sf} = 0.36 \pm 0.08\,\SI{}{\micro\electronvolt}$.

Finally, we investigate the underlying mechanism responsible for the spin-flip tunnel coupling between $T_+^{(1,3)}$ and $S^{(0,4)}$. We consider two common candidate mechanisms: (i) Hyperfine interaction between the electron spins and the nuclear spin bath can result in a spin flip, after which the system can transition to (0,4) via a spin-conserving tunneling process. This yields an effective coupling energy of $t_{\rm sf} \sim \frac{1}{\sqrt 2} t (K / |B_\parallel|)$, where $K$ is the typical magnitude of the effective nuclear fields acting on the electrons in the dots. (ii) Spin-orbit coupling can induce a rotation of the spin state of electrons ``during'' tunneling between the dots, resulting in a coherent coupling between $T_+^{(1,3)}$ and $S^{(0,4)}$ of the order $t_{\rm sf} \sim \frac{1}{\sqrt 2} t (d / l_{\rm so})$, where $d$ is the distance between the dots and $l_{\rm so}$ the spin-orbit length along the interdot axis\,\cite{danon_spin-flip_2013}.
We see that mechanism (i) should result in a significant field dependence of $t_{\rm sf}$ and thus $I_{\rm max}$, whereas mechanism (ii) should produce a mostly field-independent $t_{\rm sf}$. In Fig.\,\ref{fig6}\,(d) we plot the fitted values of $I_{\rm max}$ as a function of $B_\parallel$, and we see that the peak height is rather constant, suggesting that the spin-orbit mechanism is dominating. Besides, for a hyperfine-driven spin-flip tunnel coupling the estimate of $K \sim \SI{0.6}{\milli\tesla}$ (see next Section) would produce $t_{\rm sf} \lesssim \SI{20}{\nano\electronvolt}$ for fields $|B_\parallel| > \SI{500}{\milli\tesla}$, which is far too small to explain the data. Thus attributing $t_{\rm sf}$ to spin-orbit coupling, we can estimate the spin-orbit length to be $l_{\rm so} = 8.6 \pm 2.4\,\SI{}{\micro\meter}$, assuming a spacing of $\SI{200}{\nano\meter}$ between the dot centers. This is consistent with all the other estimates presented above.

\begin{figure}[]
	\includegraphics[width=\columnwidth]{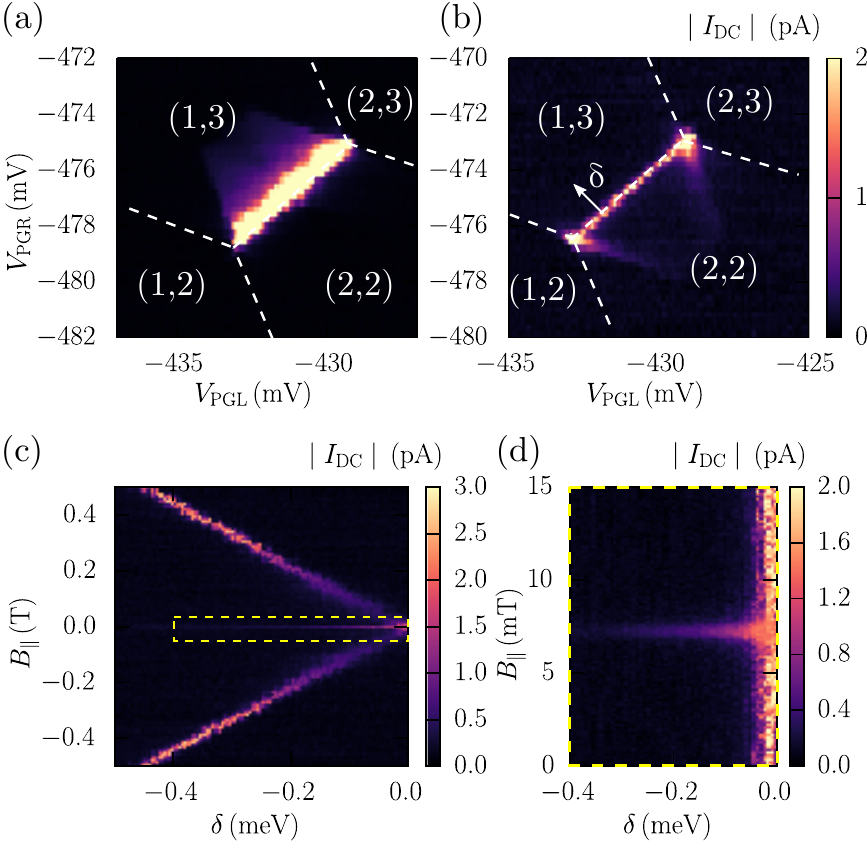}
	\caption{(a) Magnitude of the current $\vert I_{\mathrm{DC}}\vert$ through the double quantum dot system at weak interdot tunnel coupling and at a finite bias of $V_{\mathrm{DC}}=\SI{+500}{\micro\volt}$, close to the triple points involving the (2,2) to (1,3) transition. (b) Transitions from (1,3) to (2,2) at $V_{\mathrm{DC}}=\SI{-500}{\micro\volt}$ are suppressed due to singlet-triplet spin blockade. At weak interdot tunnel coupling, a leakage current is observed when the tunneling process is resonant. (c) Dependence of this leakage current on  $\delta$ and $B_{\parallel}$. A narrow resonance around zero external magnetic field is observed. (d) Zoom into the yellow box in (c) showing more clearly the zero-field resonance arising from hyperfine coupling to the nuclear spins of the host crystal. The offset from $B_{\parallel}=0$ is due to residual magnetization of our experimental setup.}  
	\label{fig7}
\end{figure}
\subsection{Spin Blockade at Weak Interdot Tunnel Coupling}
Upon reducing the tunnel coupling between the two dots, we enter a regime in which the spin blockade manifests itself in a qualitatively different way\,\cite{pfund_suppression_2007,nadj-perge_disentangling_2010,koppens_control_2005}. Fig.\,\ref{fig7}\,(a) shows the current in the region close to the boundary between the (2,2) and (1,3) charge regions at $V_{\rm DC} = \SI{500}{\micro\volt}$. The finite bias triangles in this forward direction show a pronounced baseline, corresponding to resonant interdot tunneling, and quickly decaying current levels towards the tips of the triangles. The situation at reverse bias in the spin blockade regime, can be seen in Fig.\,\ref{fig7}\,(b). Similar to the forward direction, a pronounced baseline of the triangles is visible, but inside the triangles the current is now suppressed, which is qualitatively different from the strong-coupling triangles shown in Fig.\,\ref{fig5}\,(c).
In Fig.\,\ref{fig7}\,(c) we plot the leakage current as a function of $B_\parallel$ and $\delta$, where $\delta$ is now measured along the white dashed line in Fig.\,\ref{fig7}\,(b). The strongest resonances appear in a V-shape like in Fig.\,\ref{fig6}\,(a), mapping out where the (1,3) $T_+$ ground state is resonant with the (2,2) singlet state. In addition to these resonances we also observe a very narrow resonance at zero magnetic field, in contrast to the zero-field dip we found in the strong-coupling regime. This peak in the leakage current is caused by hyperfine coupling between the electron spins and the nuclear spins in the surrounding crystal\,\cite{ono_nuclear-spin-induced_2004,koppens_control_2005,johnson_tripletsinglet_2005}. This coupling is now stronger than the exchange effects caused by the interdot tunnel coupling and can thus efficiently mix all four (1,3) spin states at small magnetic fields, thereby lifting the blockade of the triplet states. From the zoom of this resonance in Fig.\,\ref{fig7}\,(d), we find a full width at half maximum of $\approx\SI{0.6}{\milli\tesla}$; this indicates the typical magnitude of the random effective nuclear fields acting on the electron spins. This value is consistent with other values reported for InAs-based quantum dots\,\cite{pfund_suppression_2007,nadj-perge_disentangling_2010,nadj-perge_spinorbit_2010} taking into account our comparatively large dot volume and the large $g$-factor in our system.

\section{Conclusion}
In conclusion, we have presented the first realization of high-quality gate-defined few-electron single and double quantum dots in an InAs two-dimensional electron gas. 
We obtained DC control over our dots with similar precision as is common for GaAs- and Si-based dots, which have been around for decades.
In the single quantum dot, we observed a pronounced Kondo effect of the last electron and we characterized this effect consistently with theory and previous experiments in other materials.
Additionally, we performed excited state spectroscopy which allowed us to determine the $g$-factor $g \approx 16$ and put a lower bound on the effective spin-orbit length in the system.
In the double quantum dot we observed Pauli spin blockade, and analyzed the leakage current for both strong and weak interdot tunnel coupling to arrive at estimates for the typical magnitude of the random hyperfine fields on the dots $K \sim \SI{0.6}{\milli\tesla}$ and the spin-orbit length $l_{\rm so} \sim \SI{8}{\micro\meter}$, which is almost two orders of magnitude longer than typically measured in InAs-based nanostructures.
Together with the low effective electronic mass in InAs, these results put InAs on the map as a competitive base material for spin qubits.

The unusually large spin-orbit length we measured is most likely due to the very symmetric design of our quantum well in the out-of-plane direction, since any small asymmetry in the potential is leveraged into strong spin-orbit coupling by the large Rashba coefficient in InAs.
This combination of having weak spin-orbit coupling in a material with a large Rashba coefficient could potentially allow for engineering two-dimensional structures where gate-induced electric fields generate strong spin-orbit interaction locally in space and/or time.

\begin{acknowledgments}
The authors acknowledge the support of the ETH FIRST laboratory and the financial support of the Swiss Science Foundation (Schweizerischer Nationalfonds, NCCR QSIT). The work at Purdue was funded by Microsoft Quantum. J.D. acknowledges support through FRIPRO-project 274853, which is funded by the Research Council of Norway.
\end{acknowledgments}

\bibliography{bibl}
\end{document}